\def\D0{D\O~}

\documentstyle[preprint,aps,floats]{revtex}
\ifx\nopictures Y \else \input epsf.tex \fi
\begin{document}
\draft
\tighten
\baselineskip=18pt

\title{%
Comment on the $W + 1$ jet to $W + 0$ jets ratio}
\author{%
C. Bal\'azs\thanks{E-mail address: balazs@pa.msu.edu} and 
C.--P. Yuan\thanks{E-mail address: yuan@pa.msu.edu}}
\date{\today}
\address{%
Department of Physics and Astronomy, Michigan State University, \\
East Lansing, MI 48824, U.S.A.}
\maketitle

\begin{flushleft}
{MSUHEP-70428 \\ CTEQ-706}
\end{flushleft}

\begin{abstract}
To offer a possible resolution to the apparent discrepancy between the 
experimental and the theoretical values of the $W + 1$ jet to $W + 0$
jets ratio reported by the \D0 group,
we examine the effects of the multiple soft gluon radiation on
the $W$ boson production at the Tevatron. Based on the calculation of the $W$ 
boson transverse momentum ($Q_T$) distribution in the Collins-Soper-Sterman
resummation formalism, we conclude that the effect of the soft gluon radiation 
is important in the region of $Q_T < 50$~GeV, and it can be better tested by 
a more inclusive observable $R_{CSS}(Q_T^{\min})\equiv
\frac{\sigma (Q_T>Q_T^{\min })}{\sigma _{Total}}$. 
\end{abstract}

Reports on the $W + 1$ jet to $W + 0$ jets ratio by the Fermilab \D0
collaboration \cite{D0alphaS,D0R10} have repeatedly shown a
discrepancy between the experimental data 
and the next-to-leading-order (NLO) QCD predictions of the 
DYRAD Monte Carlo program \cite{DYRAD}.
In accordance with the experimental situation, a DYRAD event is considered 
to represent $W + 0$ jets ($W + 1$ jet) event if the
transverse momentum ($E_T$) of the jet  is smaller 
(larger) than a certain $E_T^{\min}$ value.
Using the above definition, the $W + 1$ jet $E_T$ cross section is calculated 
in ${\cal O}(\alpha_S^2)$
while the $W + 0$ jets $E_T$ distribution is calculated in 
${\cal O}(\alpha_S)$~\cite{Yu}.
From these $W +$ jet cross sections, the following ratio is formed:
\begin{eqnarray}
R^{10}_{jet}(E_T^{\min}) = 
\frac
{\displaystyle \int_{E_T^{\min}}^{E_T^{\max}} d E_T \frac{d \sigma}{d E_T}}
{\displaystyle \int_0^{E_T^{\min}}            d E_T \frac{d \sigma}{d E_T}},  
\end{eqnarray}
where $E_T^{\max}$ is the maximal $E_T$ allowed by the phase space.
This ratio is then compared with the experimental results.
The DYRAD prediction of $R^{10}_{jet}$ is found to be consistently lower than 
the experimental central values by about 30\%.
Since, with the increase of $E_T^{\min}$ the 
error bars of the  experimental data increase,
the confidence level of the statistical significance of the deviation is 
smaller in the $E_T^{\min} > 50$ GeV region.
Therefore, we focus our attention on the $E_T^{\min} < 50$~GeV section, 
and ask the question: What physics can be responsible for this deviation?

The \D0 analysis offers the uncertainty of the gluon parton distribution
as the most likely reason for the discrepancy \cite{D0R10}. 
In this work we propose an additional possible explanation which originates 
from the perturbative QCD theory.
We note that the situation is very similar to the one in the direct
photon production, another process in which a vector boson produced together 
with a jet (or several jets).
The NLO prediction of the 
transverse momentum ($p_T$) distribution of the photon is
systematically lower than the measured cross sections 
\cite{Joey} for the low $p_T$ region.

Different parts of the answer to the $W +$ jet puzzle might come from 
different sorts of physics. To illustrate this, we point out that the ratio 
$R^{10}_{jet}$ has several shortcomings from a theoretical point of view:
\begin{quote}
$\bullet$ The NLO calculation of $R^{10}_{jet}$ might not be
sufficient to describe the data in the low transverse momentum region
for it does not include the large effect of the multiple soft gluon
emission.
\end{quote}
\begin{quote}
$\bullet$ Calculating the numerator of $R^{10}_{jet}$ in ${\cal O}(\alpha_S^2)$
and the denominator in ${\cal O}(\alpha_S)$ implies a discontinuity in the
$E_T$ distribution at $E_T^{\min}$. It is therefore less natural than, say, a
pure ${\cal O}(\alpha_S^2)$ calculation.
\end{quote}
\begin{quote}
$\bullet$ The value of $R^{10}_{jet}$ depends on the 
detailed definition of the jet in both the 
theoretical calculation and the experimental measurement. 
\end{quote}
Each of these deficiencies may contribute to a different degree to the 
disagreement in various $E_T$ regions. 
Since no better calculation of the jet $E_T$ cross section is available
than the one used by \D0, in order to analyze the situation, we turn to
the calculation of the transverse momentum ($Q_T$) of the $W$ boson.
In ${\cal O}(\alpha_S)$, the transverse momenta of the jet and the
$W$ boson are the same: $E_T = Q_T$. 
In addition, the $Q_T$ distribution of the $W$ boson is theoretically well 
understood, and the contributions from the multiple soft gluon radiation can 
be resummed using the Collins-Soper-Sterman formalism \cite{CSS,WRes,RCSS}.

We can form the ratio $R^{10}_{W}$ using the $W$ boson transverse momentum in 
the place of the jet $E_T$:
\begin{eqnarray}
R^{10}_{W}(Q_T^{\min}) = 
\frac
{\displaystyle \int_{Q_T^{\min}}^{Q_T^{\max}} d Q_T \frac{d \sigma}{d Q_T}}
{\displaystyle \int_0^{Q_T^{\min}}            d Q_T \frac{d \sigma}{d Q_T}},  
\end{eqnarray}
where $Q_T^{\max}$ is the largest $Q_T$ allowed by the phase space.
\begin{figure*}[t]
\begin{center}
\begin{tabular}{cc}
\ifx\nopictures Y \else{ 
\epsfysize=6.0cm 
\epsffile{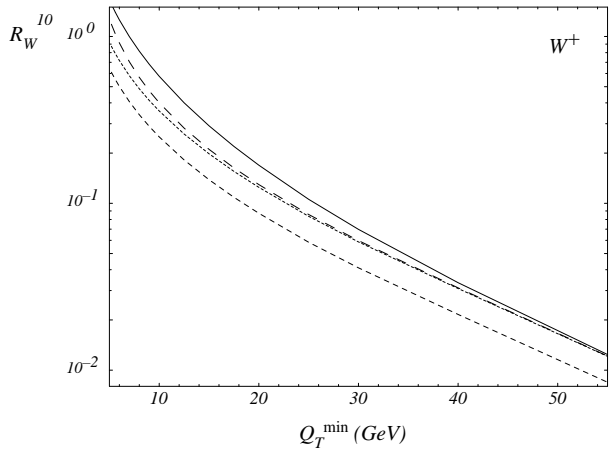} } 
\fi &  \\ 
\end{tabular}
\end{center}
\caption{The ratio $R^{10}_W$ 
calculated in the resummed formalism (solid), 
in ${\cal O}(\alpha_S^2)$ (long dash),
in ${\cal O}(\alpha_S)$ (short dash), 
and with numerator in ${\cal O}(\alpha_S^2)$ and 
denominator in ${\cal O}(\alpha_S)$ (dots).}
\label{fig:R10W}
\end{figure*}
In Fig.~\ref{fig:R10W}, we plot the ratio $R^{10}_W$ for calculations done 
in different orders of the strong coupling constant
$\alpha_S$.\footnote{We use the ResBos 
Monte Carlo code \cite{WRes}, $\sqrt{S} = 1.8~TeV$, and the CTEQ4M 
parton distribution.}
The difference of the ${\cal O}(\alpha_S)$ (short dashed) and 
${\cal O}(\alpha_S^2)$ (long dashed) curves
indicates that the K-factor is about 1.4 in the region of interest, which
suggests that higher order perturbative contributions might have to be 
considered. From comparing the resummed (solid) and the ${\cal O}(\alpha_S^2)$ 
(long dashed) curves, we infer that the effects of the multiple soft gluon 
radiation increase the $Q_T$ cross section for $Q_T < 50$~GeV.
This increase over the ${\cal O}(\alpha^2_S)$ 
rate is about 30\% around 
$Q_T = 20$~GeV, and remains sizable (more than 5\%) even at $Q_T = 40$~GeV.
The dotted curve, which is calculated with the mismatched numerator
(to ${\cal O}(\alpha_S^2)$) and denominator (to ${\cal O}(\alpha_S)$), 
runs under the ${\cal O}(\alpha_S^2)$ (long dashed) curve, 
but the difference coming from this mismatch is small in the 
$Q_T > 20$~GeV region.

Based on the results obtained for the $Q_T$ distribution of the $W$
boson, we conclude that soft gluon effects are important in the $W +$ jet 
production in the region of $Q_T < 50$~GeV. 
The ratio $R_{CSS} \equiv 1/(1 + R^{10}_W)$ \cite{RCSS} 
is more suitable to be compared to the experimental data for 
it is a more inclusive observable which does not
involve any jet measurement but includes 
the large effect of multiple soft gluon
contribution to all orders in $\alpha_S$.

We are grateful to R. Brock, T. Joffe-Minor, W.K. Tung and 
H. Weerts for helpful discussions and suggestions. 
This work was supported in part by NSF under grant No. PHY-9507683.


\end{document}